\documentstyle[aps,prl,floats,epsf]{revtex} 

\begin{document} 
\draft  
\title {Evolution of the quasiparticle spectral function in
cuprates}  
\author{S. Misra, R. Gatt, T. Schmauder, Andrey V. Chubukov
and M. Onellion\cite{core}}    
\address{Department of Physics,
University of Wisconsin, Madison, WI 53706} 
\author{M. Zacchigna,
I. Vobornik, F. Zwick, M. Grioni and G. Margaritondo}    
\address{$\acute{E}$cole Polytechnique F$\acute{e}$d$\acute{e}$rale 
Lausanne, CH-1015 Lausanne, Switzerland} 
\author{C.  Quitmann}
\address{Experimentalphysik I, Universit$\ddot{a}$t Dortmund, D-44221
Dortmund, Germany} 
\author{C. Kendziora}  
\address{Naval Research Laboratory, Washington D.C.}   
\date{\today}  

\maketitle 

\begin{abstract}
We analyzed photoemssion data for several doping levels of the
$Bi_2Sr_2CaCu_2O_{8+x}$  compounds, ranging from overdoped to
underdoped.  We show that the high frequency part of the spectra near
$(0,\pi)$ can be described by Fermi liquid theory in the overdoped
regime, but exhibits a non-Fermi liquid behavior in the underdoped
regime. We further demonstrate that this novel behavior fits
reasonably well to a $1/\sqrt{\omega}$ behavior suggested for systems
with strong spin fluctuations.
\end{abstract}

One of the most intriguing characteristics of the  cuprates is their
qualitative change in electronic properties  with doping. These
changes have been attributed mostly to the pseudogap that appears in
the  normal state at frequencies comparable to the superconducting
gap.  The goal of this  communication is to show that in addition to
the pseudogap, the high frequency part of photoemission spectra
exhibit qualitative changes with decreasing carrier concentration. We
argue below that this evolution with doping indicates that the
cuprates are near a quantum critical point and exhibit a non-Fermi
liquid spectral function for high frequencies.

A series of recent papers have reported that the spectral weight of
the quasiparticle peak near the Fermi surface becomes smaller for
underdoped samples, and that there is a transfer of spectral weight to
the high-frequency part of the spectrum \cite{H}. In this paper, we
present a functional analysis of such data. Our main result is that
the photoemission intensity $I(\omega)$ undergoes a particular type of
crossover behavior with decreasing doping (since energy is
$\hbar\omega$, we subsequently refer to the frequency $\omega$).  The
spectral function changes  from a behavior which can be reasonably
well described by the Fermi liquid form $I^{-1} (\omega) \propto
(\Gamma^2 + \omega^2)$ to a behavior which to good accuracy can be
described by a non-Fermi liquid form $I^{-1}(\omega) \propto \sqrt
\omega$.  Such behavior is predicted when the dominant effect of
quasiparticle decay is the interaction with overdamped spin
fluctuations \cite{I} .

We begin by  reviewing the theoretical aspects of the problem.  In the
sudden  approximation, the photoemission intensity at a given momentum
$k$ is given by $I_k(\omega) = A_k (\omega) n(\omega)$ where
$A_k(\omega) = (1/\pi) Im G_k(\omega)$ is the quasiparticle spectral
function, and $n(\omega)$ is the Fermi function which selects states
occupied by  electrons.  Fermi liquid theory predicts that, regardless
of how strong the interaction is, the single-particle Green's function
at low frequencies  has the form 
\begin{equation}
G_k (\omega) = \frac{Z}{(\omega - (\epsilon_k - \epsilon_F) + i \omega
|\omega|/\Gamma)},
\label{FL}
\end{equation}
 where $\epsilon_F$ is the Fermi energy.  For momenta near the Fermi
surace, i.e.,  at small $\epsilon_k - \epsilon_F$, the spectral
function which emerges from $G_k (\omega)$ possesses a quasiparticle
peak at $\omega = \epsilon_k - \epsilon_F$; the width of the peak
decreases as $(\epsilon_k - \epsilon_F)^2$ with the approach of the
Fermi surface. At the Fermi-surface, the quasiparticle peak transforms
into a $\delta$-function for $\omega =0$, while for frequencies
$\omega \leq \Gamma$, $A_k (\omega)$ takes the form $A^{-1}_k (\omega)
\propto (\omega^2 + \Gamma^2)$.  The inverse proportionality to
$\omega^2$ at low frequencies is a fundamental property of Fermi
liquids which does not depend on details of the quasiparticle
interaction. However, $\Gamma$, which serves as  the upper frequency
cutoff for this universal behavior, is model dependent.  When the
Fermi system is far from any instability, $\Gamma$ is of the order as
$\epsilon_F$, which is typically the upper cutoff for a low energy
description (i.e., expansion in powers of $\omega$).  This changes if
the system is close to a quantum critical point where it undergoes a
spontaneous symmetry breaking. In this situation, $\Gamma \ll
\epsilon_F$ and $\Gamma$ scales with the distance from the critical
point. For $\omega \ll \Gamma$, the system still possesses Fermi
liquid behavior- i.e., $A^{-1}_k (\omega) \propto (\omega^2 +
\Gamma^2)$. However, for $\Gamma \ll \omega \ll \epsilon_F$ the system
crosses over into the region which is in the basin of attraction of
the quantum critical point. In this region the system behavior is
governed by critical fluctuations.  These fluctuations impose a novel
frequency dependence of the spectral function up to frequencies
comparable to the Fermi energy. It is essential that this picture
survives even if the system trajectory in the phase space of
parameters slightly bypasses the quantum critical point (e.g., the
actual transition into the symmery breaking state can be a weak first
order transition or involve an intermediate phase) or the system
possesses a low-energy crossover from one type of critical behavior to
that associated with small but relevant perturbations around the
quantum critical point (in cuprates this may be due to a Fermi surface
evolution very near the magnetic instability). These effects reduce
the range of Fermi-liquid behavior to even smaller frequencies, but
they do not modify the behavior at $\omega > \Gamma$ as long as
$\Gamma$ remains the largest energy scale associated with instability.

In recent years there have been a number of suggestions for possible
quantum critical points in the cuprates \cite{crit}. In our opinion,
the most plausable candidate is the instability point towards
antiferromagnetism.  The system approaches this critical point as the
carrier concentration decreases, and the system is antiferromagnetic
at half-filling. The quantitative measure of closeness to a magnetic
transition is the spin correlation length, $\xi$. When $\xi$ is large
enough, the system behavior is predominantly critical over a wide
frequency range. As $\xi$ clearly increases with decreasing doping,
the range of critical behavior also increases. 

Recently, one of us (AC) considered the interaction between fermions
and overdamped spin fluctuations and found that  $\Gamma \sim
\omega_{sf}$ where $\omega_{sf} \propto \xi^{-2}$ is a typical
spin-fluctuation scale \cite{I}. More specifically, the fermionic
Green's function along some portion of the Fermi surface near
$(0,\pi)$ and related points was shown to have the form
\begin{equation}
G^{-1}(\omega) \propto \frac{1}{Z} ~\frac{2\omega}{1 + \sqrt{1 -
i|\omega|/\omega_{sf}}},
\label{G}
\end{equation}
where $Z \propto \xi^{-1}$.  At small frequencies, $\omega \leq
\omega_{sf}$, this Green's function has a conventional Fermi-liquid
form  $G^{-1} (\omega) \propto (\omega + i \omega
|\omega|/(4\omega_{sf}))$.  At higher frequencies, however, the
Fermi-liquid form crosses over to a region of novel frequency
dependence- $G (\omega) \propto A e^{-i\pi/4} |\omega|^{-1/2}  sgn
\omega$, where $A = Z\omega_{sf}^{-1/2}$. As $A$  is independent of
$\xi$, it remains finite when $\xi \rightarrow \infty$, as it should
for quantum critical behavior. One can probe this behavior at finite
$\xi$ for $|\omega| \gg \omega_{sf}$. Values of $\omega_{sf}$ for
various dopings have been extracted from available NMR data
\cite{pines}.  The NMR data indicate that $\omega_{sf}$ (at a given T)
increases with doping and remains $\sim 20-30 meV$  (i.e., small
compared to $\epsilon_F$) for overdoped cuprates. From this
perspective,   some features of the crossover to $\sqrt{\omega}$
behavior should be observed even in the data for overdoped cuprates.

The tendency towards quantum critical behavior is much less pronounced
 near the Fermi surface crossing along the zone diagonal
 ($\langle\pi,\pi\rangle$ direction).  Here, calculations show that
 the analog of $\omega_{sf}$ remains finite for $\xi \rightarrow
 \infty$ with a value comparable to $\epsilon_F$.  The system is
 therefore in the moderate coupling regime; $A(\omega)$ at
 intermediate $\omega$ should display a frequency dependence which is
 intermediate beween quantum-critical and Fermi liquid forms.
 
We now turn to the analysis of the photoemission data.  The
photoemission data were measured using angle-resolved photoemission
equipment at the Wisconsin Synchrotron Radiation Center and at
$\rm\acute{E}$cole Polytechnique F$\rm\acute{e}$d$\rm\acute{e}$rale
Lausanne; the equipment details are provided in Ref. \cite{J}.
Samples were fabricated as reported in the literature \cite{K}, and
were transferred via a load-lock system so as to retain the oxygen
content.  The data were taken at a temperature well into the normal
state (typically $200K$ for underdoped samples, $100K$ for overdoped
samples), and above the temperature, $T^*$, below which a pseudogap
would be observed.  In addition, we analyzed the data from
Ref. \cite{L} and found our conclusions to be consistent with that
report.  Altogether, we analyzed data on seven doping levels,
including overdoped ($T_C \approx 52K$, $65K$ and $75K$), optimally
doped ($T_C \approx 90K$), and underdoped ($T_C \approx 60K$ and
$30K$), in addition to analyzing underdoped data ($T_C \approx 85K$)
from Ref. \cite{L}. The details of all samples and doping levels are
reported elsewhere \cite{M}.
 
We analyzed the photoemission data using seven different techniques,
including fitting using the chi-squared criterion (with four
different background subtraction methods), the Kolmogrov-Smirnov
criterion, the ratio of data at different binding energies, and the
inverse of the data.  We provide a complete, detailed discussion of
these analysis methods elsewhere \cite{M}. All seven methods led to
the same conclusion, that there is a qualitative change in the nature
of the spectral function with doping. Due to space limitations, we
emphasize two of these methods in this report. 

We first summarize our results.  For overdoped cuprates near
$(0,\pi)$, we found that the data below $\sim 250 meV$ (the exact
value depends somewhat on the background subtraction method employed)
were consistent with the Fermi liquid Green's function (Eq. \ref{FL})
and were inconsistent with the Green's function of Eq. \ref{G}. For
higher frequencies, however, we found some evidence for a crossover to
quantum-critical behavior.  For overdoped cuprates along the
$\langle\pi,\pi\rangle$ direction, the data were also consistent with
Eq. \ref{FL}.  For underdoped samples near $(0,\pi)$, we found that
the data were consistent with Eq. \ref{G}, and were inconsistent with
a Fermi liquid Green's function  over the entire frequency range
studied.  In the $\langle\pi,\pi\rangle$ direction, the data are fit
equally well by either the Fermi liquid or the non-Fermi-liquid
spectral functions.

We now discuss in more detail how we actually fitted the data.  We
began by checking whether the  finite energy resolution of our
equipment significantly modified the functional forms of either
Eqs. \ref{FL} or \ref{G}. To do this, we generated spectra by
convoluting  each of the two theoretical spectral functions with the
experimental resolution function, and subsequently fitted the results
using the unbroadened functional forms of Eqs. \ref{FL} and \ref{G}.
We found that the finite energy resolution of the electron analyzer
has only a small impact on the analyses and  could be almost fully
absorbed into small renormalization of the input parameters,
$\omega_{sf}$ and $\Gamma$.

\begin{figure}[p]

\leavevmode 
\epsfclipon
\centerline{\hbox{ 
\epsfxsize=7.0in{\epsffile{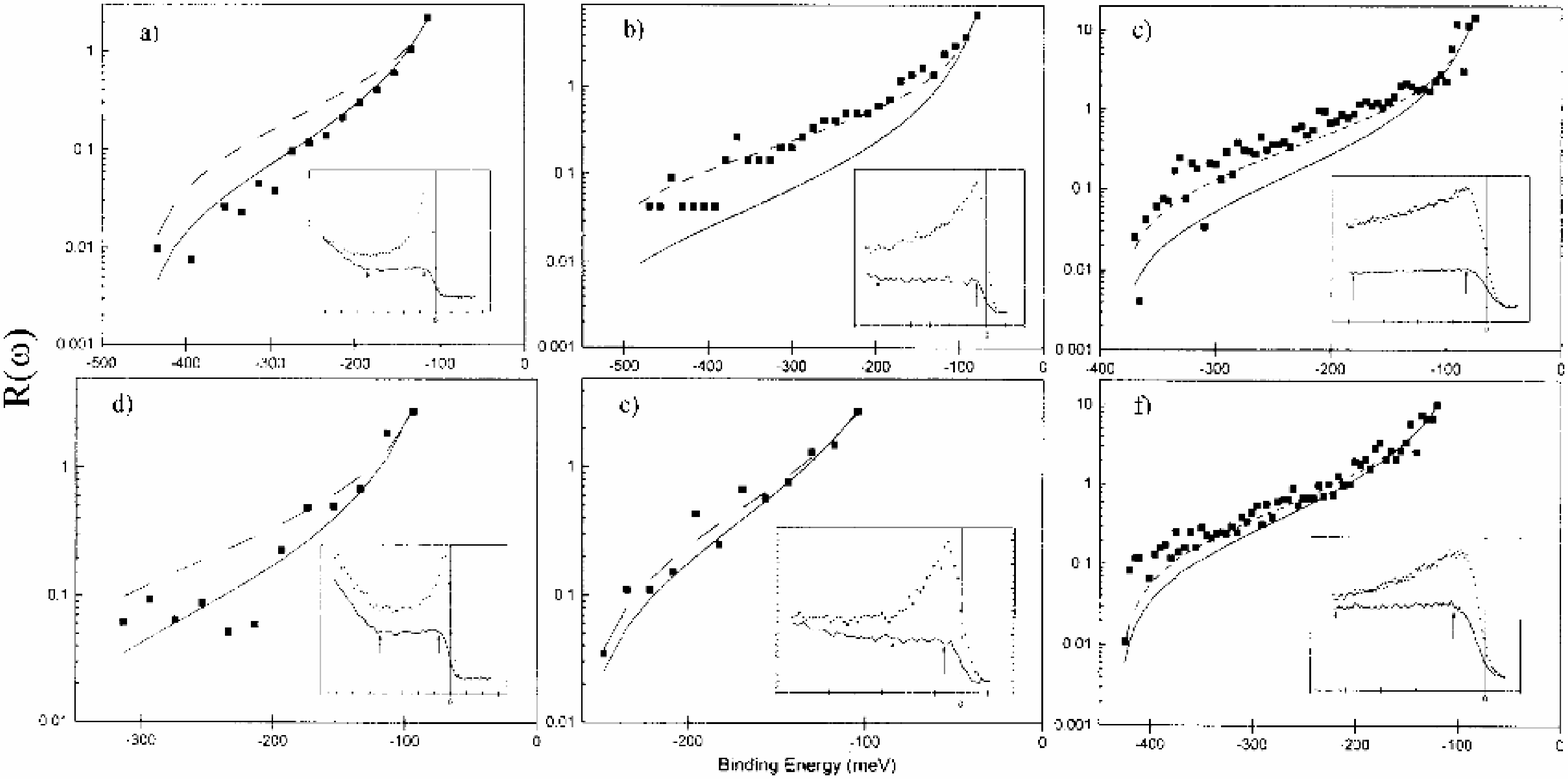}} 
}}

\caption{
  The fits of the normalized photoemission intensity $R(\omega)$  by
  the Fermi-liquid frequency dependence, Eq. \protect\ref{FL} (dotted
  line), and by the spin-fluctuation form, Eq. \protect\ref{G} (solid
  line). The horizontal axis is the electron binding energy from 0
  (the Fermi energy) and the hash marks 100 $meV$ apart. The vertical
  axis is $R(\omega) $ on a logarithmic scale.  Figs. (a)-(c) are fits
  to the data near the $(0,\pi)$ point  including: (a) oxygen
  underdoped, $T_C \approx 60K$ material, (b) Dy-underdoped,  $T_C
  \approx 85K$ material from Ref.  \protect\cite{L}, and (c) oxygen
  overdoped, $T_C \approx 65K$ material.  Figs (d)-(f) are fits to the
  data near the Fermi surface crossing in the $\langle\pi,\pi\rangle$
  direction, including: (d) oxygen underdoped, $T_C \approx 30K$
  material, (e) Dy- underdoped, $T_C \approx 85K$ from
  Ref. \protect\cite{L} (same as in Fig. (b)), and (f) oxygen
  overdoped, $T_C \approx 65K$ material.  Insets to the figures show
  angle-resolved photoemission spectra taken near the Fermi surface
  (upper spectra) and far away from the Fermi surface (lower
  spectra). The horizontal axis is the electron binding energy, from 0
  (Fermi energy) and the hash marks are 100 $meV$ apart.  The arrows
  indicate the frequency range for which the spectra far away from the
  Fermi surface are frequency independent. This is the range over
  which we fitted $R(\omega)$.}
\label{fig2}
\end{figure}

In our first method, we took special care in dealing with the background.
 The key assumption we made is that photoemission spectra for
 wavevectors far from the Fermi surface are indistinguishable from the
 background.  Analyzing the data in these $k$-ranges, we observe that
 they are nearly frequency independent over a relatively wide
 frequency range.  The insets of Fig. \ref{fig2}  illustrate this
 situation (the flat regions are indicated by arrows). We further
 assumed that the flat background is independent of $k$, i.e., it
 remains the same near the Fermi surface.  We then eliminate this
 background by replotting the data in the form $R(\omega) =
 ((I(\omega) - I(\omega_{max}))/(I(\omega_{min}) - I(\omega))$ where
 $\omega_{min}$ and $\omega_{max}$ are the lower and upper boundaries
 of the region where the background is independent of frequency.
 Plotting the data in this form additionally allows us to eliminate
 the overall factor in the intensity.  Finally, we fit $R(\omega)$ by
 Fermi-liquid and spin-fluctuation forms using $\omega_{sf}$ and
 $\Gamma$ as adjustable parameters and check which form works better. 

Applying this method to experimental data, we obtain the results
presented in Fig. \ref{fig2}.  The actual  spectra from which the fits
were obtained are the upper curves in the insets of each part of
Fig. \ref{fig2}.  We include three sets of data- oxygen overdoped
material  ($T_C \approx 65K$),  oxygen underdoped material ($T_C
\approx 30K$ and $60K$), and Dy-doped underdoped material ($T_C
\approx 85K$) from Ref. \cite{L}. The spectra analyzed are those for
the $k$-points closest to the Fermi
surface. For overdoped samples near the $(0,\pi)$ point, the data
consistently yield a better fit for Eq. \ref{FL}.  Along the
$\langle\pi,\pi\rangle$ direction, the data are also fit better by
Eq. \ref{FL}, although the error bars are more significant. 
For underdoped materials near the $(0,\pi)$ point, the spectral
function from Eq. \ref{G} yields a much better fit than that of a
Fermi liquid.  Furthermore, we consistently obtain a spin fluctuation
energy of 30- 40 $meV$, which is a bit larger than, but still
comparable to, the $\omega_{sf} \sim 10-20 meV$ extracted from NMR.  Along
the $\langle\pi,\pi\rangle$ direction, both spectral functions fit the
data equally well, thus we cannot judge which form works better.

\begin{figure}[p]

\leavevmode 
\epsfclipon 
\centerline{\hbox{
\epsfxsize=3.4in{\epsffile{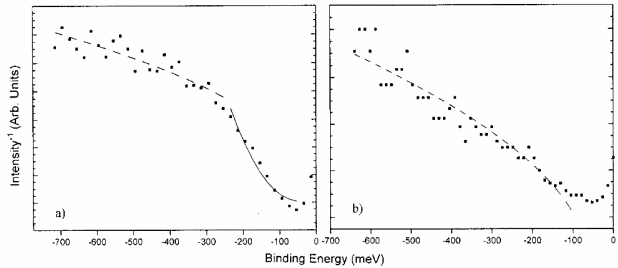}}
}}

\caption{Results of the second analysis method of fitting the data. 
We substracted the data far away from the Fermi surface (which we
identify with the  background) from the data taken near the Fermi
surface crossing near the $(0,\pi)$ point for all $k-$values and
inverted the difference . We then fitted the inverse  intensity by the
Fermi-liquid, $\omega^2$ form, (dotted line) and the $\sqrt\omega$
form (solid line),  which is the high-frequency limit of the
spin-fluctuation form (Eq. \protect \ref{G}).  Figs (a) and (b) are
the fits for  oxygen overdoped, $T_C \approx 65K$ material and  oxygen
underdoped, $T_C \approx 60K$ material, respectively.  Observe that
the $\omega^2$ form does not fit the data for underdoped sample at
all. For overdoped sample, the $\omega^2$ form fits data for binding
energy up to $\sim 250meV$, above which there is a crossover to the
$\sqrt\omega$ dependence.}
\label{fig3}
\end{figure}

In the second  method of analysis, we assumed that the background is
 independent on $k$ over the whole frequency range studied in
 photoemission. After making this assumption, we
 subtracted the data for $k$-values far away from the Fermi surface
 from the data near the $(0,\pi)$ point, inverted the subtracted
 data, and fitted the results to both Fermi liquid ($\omega^2$)
 and $\sqrt\omega$ forms. This method allowed us to fit the data over a
 wider frequency range than the first one. The results are shown in
 Fig. \ref{fig3}. The results provide additional insights to the
 results in Fig. \ref{fig2}.  For overdoped samples, we find that the
 data are fit best by a quadratic (Fermi liquid) form for binding
 energies up to $\sim 250 meV$, and by a $\sqrt\omega$ form  for
 higher binding energies. The exact binding energy of the crossover
 depends somewhat on the background subtraction method employed.  For
 underdoped samples, the data in Fig. \ref{fig3}  are better fit by a
 $\sqrt\omega$ form for the entire binding energy range above $\sim
 100 meV$.  We also performed the same analysis for the data taken
 near the Fermi surface crossing along the $\langle\pi,\pi\rangle$
 direction (not shown in Fig. \ref{fig3}). For both overdoped and
 underdoped samples, the Fermi liquid and non-Fermi liquid forms fit 
 the data equally well. As a consequence,
 it is difficult to assess which form works better in the along the
 $\langle\pi,\pi\rangle$ direction for any doping level.

In summary, theory  predicts that when doping decreases, the
Fermi-liquid form of the spectral function near $(0,\pi)$ exists in
progressively smaller range of low frequencies, while at higher
frequencies the spectral function crosses over to a novel  frequency
dependence, $A^{-1}(\omega) \propto \sqrt{\omega}$, associated with
the closeness to an antiferromagnetic quantum critical
point. Experimentally, we have measured and analyzed the high
frequency part of the photoemission data for BSCCO-2212 materials and
its variation with doping. We found that the data near $(0,\pi)$ for
underdoped cuprates agrees well with the spin-fluctuation form of the
spectral function, Eq. \ref{G}. For the frequency range we used, it
yields a non-Fermi liquid form $A^{-1}(\omega) \propto
\sqrt{|\omega|}$. In contrast, the data for overdoped cuprates is
well-fitted by the Fermi liquid form $A^{-1}(\omega) \propto (\omega^2
+ \Gamma^2)$ over a substantial frequency range (up to $\sim 250
meV$). At larger frequencies, we found some evidence for crossover to
$1/\sqrt{|\omega|}$ behavior.  Overdoped cuprates in the
$\langle\pi,\pi\rangle$ direction exhibit Fermi liquid
behavior. Neither analysis method yields conclusive results for
underdoped cuprates in the $\langle\pi,\pi\rangle$ direction.

Financial support was provided by the NSF DMR-9629839 (AC),
DMR-9632527 (SM, RG, MO), ONR through NRL (CK),  Fonds National Suisse
(EPFL, TS), and Deutsche Forschungsgemeinschaft (CQ). (SM)
acknowledges support from a Hilldale Fellowship.

\begin{references}

\bibitem[*]{core} corresponding author: M. Onellion\\  email:
onellion@comb.physics.wisc.edu

\bibitem{H} H. Ding {\it et al.}, Nature {\bf 382}, 51 (1996);
 Z.-X. Shen and J.R. Schrieffer, Phys. Rev. Lett. {\bf. 78}, 1771
 (1997); M. Norman {\it et al}, Nature {\bf 392}, 157 (1998).
 
\bibitem{I} A. Chubukov, Phys. Rev. B {\bf 52}, R3840 (1995) and
cond-mat/9709221; A.Millis, Phys. Rev. B {\bf 45}, 13047 (1995);
A.V. Chubukov and J. Schmalian, Phys. Rev. B to appear.  

\bibitem{crit}  A. Chubukov, S. Sachdev and J. Ye, Phys. Rev. B {\bf
49}, 11919 (1994);  R.  Laughlin, Phys. Rev. Lett. {\bf 79}, 1726
(1997) and unpublished; S.C. Zhang, Science {\bf 285}, 1089 (1997);
A.V. Chubukov, D.K. Morr and K. Shakhnovich, Phil. Mag. B {\bf 74},
563 (1996).

\bibitem{pines} V. Barzykin and D.  Pines, ibid, {\bf 52}, 13585
(1995). 

\bibitem{J} Jian Ma {\it et al.}, Phys. Rev. B {\bf 51} 3832, 9271
(1995).  I. Vobornik {\it et al.}, Surf. Sci., in press (1998).

\bibitem{K} C. Kendziora {\it et al.}, Physica C {\bf 257}, 74 (1996).

\bibitem{L} D.S. Marshall {\it et al.}, Phys. Rev. Lett. {\bf 76},
4841 (1996).

\bibitem{M} S. Misra {\it et al.}, unpublished.

\end{references}
\end{document}